\documentclass[twocolumn,prm,preprintnumbers,amsmath,amssymb,citeautoscript]{revtex4}
\usepackage{makecell}
\usepackage{graphicx}
\usepackage{dcolumn}
\usepackage{bm}
\usepackage{epstopdf}

\usepackage{color,hyperref}
\definecolor{blue}{rgb}{0.0,0.0,1}
\hypersetup{colorlinks,breaklinks,
            linkcolor=blue,urlcolor=blue,
            anchorcolor=blue,citecolor=blue}

\begin{document}

\preprint{MC}

\title{Crystallographic defects in Weyl semimetal LaAlGe}

\author{Inseo Kim$^{1}$}
\author{Byungkyun Kang$^{2}$}
\author{Hyunsoo Kim$^{3}$}
\author{Minseok Choi$^{1}$}
\email{minseok.choi@inha.ac.kr}
\affiliation{$^1$Department of Physics, Inha University, Incheon 22212, Korea}
\affiliation{$^2$College of Arts and Sciences, University of Delaware, Newark, Delaware 19716, United States}
\affiliation{$^3$Department of Physics, Missouri University of Science and Technology, Rolla, 65409, MO, United States}

%\author{Inseo Kim}
%\affiliation{Department of Physics, Inha University, Incheon 22212, Korea}
%\author{Byungkyun Kang}
%\affiliation{College of Arts and Sciences, University of Delaware, Newark, Delaware 19716, United States}
%\author{Hyunsoo Kim}
%\affiliation{Department of Physics, Missouri University of Science and Technology, Rolla, 65409, MO, United States}
%\author{Minseok Choi}
%\affiliation{Department of Physics, Inha University, Incheon 22212, Korea}
\date{\today}

\begin{abstract}
Crystallographic defects in a topological semimetal can result in charge doping, and the scattering due to the defects may mask its exotic transport properties.
Here, we investigate the possible crystallographic defects including vacancy and antisite in Weyl semimetal LaAlGe using hybrid-density-functional theory calculations.
We show that a considerable concentration of Al- and Ge-related defects naturally form during growth due to their low formation enthalpy. Specifically, Al can be easily replaced by Ge in the $I4_1md$ phase of LaAlGe, forming the Ge-on-Al antisite, Ge$_{\rm Al}$.
The counterpart, Al-on-Ge (Al$_{\rm Ge}$), is also probable.
The most abundant defect Ge$_{\rm Al}$ is donor-like, effectively electron-doping, and these defects are therefore not only scattering centers in the electronic transport process but may also induce the substantial vertical shift of the chemical potential. 
The results imply that the naturally occurring defects hinder both spectroscopic and transport features arising from the Weyl physics in LaAlGe.
Our work can be applied to the $R$AlGe family ($R$=rare earth) and help improve the quality of single-crystal magnetic Weyl semimetal.

\end{abstract}

%\pacs{61.72.J-, 61.72.Bb, 71.55.Ht}

\maketitle

\section{Introduction}

\textcolor{red}{}

Solid states with relativistic low-energy excitations arising from the linear band crossing are promising platforms for next-generation device technology \cite{Yang2016,Wang2023}. 
The first kind is the Dirac fermion emerging when time-reversal and parity symmetries are preserved \cite{Armitage2018}, and the two-dimensional Dirac fermion is realized at the boundary of a topological insulator \cite{Hasan2010} and in graphene \cite{Geim2007}. When, at least, one of the aforementioned symmetries is broken, and the spin-degeneracy is lifted, the Weyl fermion emerges \cite{Xu2015}. When the band intersection is sufficiently close to the chemical potential, the Weyl semimetal (WSM) exhibits unique properties, including the Veselago effect \cite{Hills2017} and selective interfacial transmission \cite{Yesilyurt2016,Hou2019}, that allow unprecedented applications \cite{Guo2023}.

%Several compounds have been identified as WSM (see Ref. \cite{Jia2016,Yan2017} for review). The first family of compounds includes the transition metal ($T$)-pnictide ($Pn$) binary alloys, $TPn$ ($T$=Na,Ta, $Pn$=P,As) \cite{Lv2015}. 

In the known WSMs, the position of the linear crossings, the Weyl nodes, is as close as a few meV to the chemical potential \cite{Xu2015,Lv2015}. In the three-dimensional system, the bulk Weyl nodes exist as a pair with opposite chirality, and they have topologically distinct characters with opposite Chern numbers \cite{Armitage2018}. The Weyl nodes are responsible for negative magnetoresistance due to chiral anomaly when the magnetic field is applied in the direction of the electrical current \cite{Ong2021}. While the Weyl nodes are separated in bulk, they are connected through the one-dimensional surface state, the Fermi arc \cite{Xu2015}. Experimental observations of chiral anomaly and the Fermi arc are regarded as direct evidence of the WSM phase.

%Recently, a new family of WSM, $Ln$Al$X$ ($Ln$=lanthanide, $X$=Ge,Si) in the LaAlSi phase lacking inversion symmetry, has attracted considerable attention since the Fermi arc was observed in LaAlGe \cite{Xu2017}. Pressure induced superconductivity in LaAlX (X=Ge,Si) - noncentrosymmetric topological superconductor \cite{Cao2022}. Compounds with $Ln$=Ce, Nd, and Pr ferromagnetically order at low temperatures, breaking time-reversal symmetry as well. The Weyl fermion in the spin polarized host would be not only scientifically intriguing but also potential new applications. 

In LaAlGe, one of the WSM family $Ln$Al$X$ ($Ln$ = lanthanide, $X$ = Ge, Si),  the angle-resolved photoemission spectroscopy (ARPES) measurement successfully resolved the Fermi arc electron states on the surface \cite{Xu2017}. However, neither anomalous Hall effect nor chiral anomaly was observed to date \cite{Hodovanets2018,Hu2020}. Quantum oscillations were not observed in a typical laboratory field range ($<$10 T) while the other members of the family ($Ln$ = Ce, Nd, and Pr) exhibit quantum oscillations at much lower fields, which is counterintuitive considering the presence of excess inelastic scattering in $Ln$Al$X$ due to localized moments. The normal state transport property in LaAlGe is consistent with conventional metal, but the residual resistivity is considerably high, which is comparable with that of a magnetic counterpart CeAlGe \cite{Hodovanets2018}. These experimental observations can be attributed to the relatively high content of native defects in LaAlGe. 

Moreover, native crystallographic defects can be a charge dopant \cite{Yu2017}, and a significant amount of such defects can shift the chemical potential.
In topological semimetals, this issue may result in a significant discrepancy between the bulk and surface chemical potentials \cite{Kim2018,Kim2022}. 
In the WSMs, the energy level of Weyl nodes relative to the chemical potential is crucial for Weyl physics, and it can be irrelevant if the nodes are far away from the chemical potential.

The LaAlGe single crystals are typically grown out of Al self-flux because of its low melting point \cite{Hodovanets2018,Hu2020}.
The volatile nature of Al may play a role and result in various crystallographic defects such as Al vacancy and subsequent interstitial or antisite associated with La and Ge that may occur during the crystal growth. Therefore, it is crucial to understand the defect energetics at the growth level.
Our aim here is to theoretically identify the potential defects that affect crystal quality and the physics of Weyl fermion in LaAlGe.

In the present work, we performed density functional theory calculation with a hybrid functional, an approach that has been demonstrated to result in accurate band structures and enthalpy of formation and has provided a reliable description of defect formation energies and defect levels in materials  \cite{Audrius_GaN_ZnO_12,AA_JAP_2016}. Details of the calculations are provided in Sec.~\ref{sec:method}.  The electronic and structural properties of LaAlGe are described in Sec.~\ref{sec:natdef} A, and electronic and structural properties of native defects are addressed in Sec.~\ref{sec:natdef} B and C. The impact of potential defects in the Weyl physics of LaAlGe is discussed in Sec.~\ref{sec:natdef} D.

%%%%%%%%%%%%%%%%%%%%%%%%%%%%%%%%%%%%%%%%%%%%%%%%%%%%%%
%%%%%%%%%%%%%%%%%%%%%%%%%%%%%%%%%%%%%%%%%%%%%%%%%%%%%%
\section{Computational approach}
\label{sec:method}

%\subsection{Density functional theory}

The calculations were based on density functional theory and the screened hybrid functional of Heyd-Scuseria-Ernzerhof (HSE) \cite{heyd:8207,krukau:224106}, implemented with the projector augmented-wave method \cite{paw} in the {\sc vasp} code \cite{vasp}.  The mixing parameter for the nonlocal Fock-exchange was set to 32\%. The tetragonal unitcell containing 12 atoms was considered for intrinsic property. The wavefunctions were expanded in a plane-wave basis set with an energy cutoff of 360 eV, and integrations over the Brillouin zone were carried out using $6\times6 \times6$ $k$-mesh.  Atomic positions were relaxed until the Hellmann-Feynman forces were reduced to less than 0.005 eV/\AA. The effect of spin-orbit coupling was also considered to look at the electronic structure of LaAlGe (See Fig. 1 in the supplementary material~\cite{supp}). However, it does not alter the defect properties (e.g. Sec.~\ref{sec:natdef} D), so we will mostly show results obtained without the spin-orbit coupling here.

The validity of the HSE electronic structure of pristine LaAlGe was checked using the $ab$-$initio$ linearized quasi-particle self-consistent GW (LQSGW) method combined with dynamical mean field theory (DMFT)~\cite{tomczak2015qsgw,choi2016first,choi2019comdmft}, implemented in the  {\sc ComDMFT} code \cite{choi2019comdmft} for DMFT and the FlapwMBPT code \cite{kutepov2017linearized} for LQSGW part. The electronic structure of the system was evaluated using LQSGW calculations~\cite{kutepov2012electronic,kutepov2017linearized}, followed by diagrammatic DMFT corrections to the local part of the GW self-energy~\cite{georges1996dynamical,metzner1989correlated,georges1992hubbard}. 
An explicit calculation of the double-counting energy and Coulomb interaction tensor was performed.
Local self-energies for La 4$f$ and La 5$d$ were obtained by solving two different single impurity models, with spin-orbital coupling included in all calculations. The electronic temperature of the system was set to 300 K.

For defect simulation, the La vacancy ($V_{\rm La}$), the Al vacancy ($V_{\rm Al}$), the Ge vacancy ($V_{\rm Ge}$), the La antisites (La$_{\rm Al}$ and La$_{\rm Ge}$), and the Al antisites (Al$_{\rm La}$ and Al$_{\rm Ge}$), and the Ge antisites (Ge$_{\rm La}$ and Ge$_{\rm Al}$) were considered. A $3\times3 \times1$ supercell containing 108 atoms was employed. The wavefunctions were expanded in a plane-wave basis set with an energy cutoff of 360 eV, and integrations over the Brillouin zone were carried out using a $2\times2 \times2$ $k$-mesh.  Atomic positions were relaxed until the Hellmann-Feynman forces were reduced to less than 0.02 eV/\AA.

%%%%%%%%%%%%%%%%%%%%%%%%%%%%%%%%%%%%%%%%%%%%%%%%%%%%%%
%\subsection{Formation energy and atomic chemical potentials}
%%%%%%%%%%%%%%%%%%%%%%%%%%%%%%%%%%%%%%%%%%%%%%%%%%%%%%

The formation energy of a defect $D$ in charge state $q$ is defined as \cite{VdW_review2004}:
\begin{eqnarray} \label{}
E^{f} (D^q) &=& E_{\rm tot}(D^q) - E_{\rm tot}({\rm LaAlGe})- \sum_{i} n_i \mu_i 
\end{eqnarray}
where $E_{\rm tot}(D^q)$ is the total energy of a supercell containing a defect $D$ in charge state $q$, and $E_{\rm tot}({\rm LaAlGe})$ is the total energy of perfect LaAlGe supercell. $n_i$ is the number of atoms of type $i$ added to ($n_i$$>$0) and/or removed from ($n_i$$<$0) the perfect crystal to form the defect, and $\mu_{i}$ ($i$ = La, Al, and Ge) are the atomic chemical potentials consisting of $\mu_i^0  + \Delta \mu_i $.
%%%%%%%%%%%%%%%%%%%%%%%%%%%%%%%%%%%%%%%%%%%%%%%%%%%%%%
%\subsection{Atomic chemical potentials}
%%%%%%%%%%%%%%%%%%%%%%%%%%%%%%%%%%%%%%%%%%%%%%%%%%%%%%
The defect formation energy depends on the atomic chemical potential $\mu_{i}$, which is taken with respect to the total energy per atom of the standard phase of the species $i$.  I.e., $\mu_{\rm La}^0$ is referenced to the total energy per atom of La bulk, $\mu_{\rm Al}^0$, that of Al bulk, and $\mu_{\rm Ge}^0$ is referenced to the total energy per atom of Ge bulk.

\begin{table}[b]
\caption{\label{enthalpy} Calculated and experimental formation enthalpies of the reference materials for atomic chemical potentials. }
\begin{ruledtabular}
\begin{tabular}{ccc}
     Materials  & Present work (eV) & Experiment (eV)\\
    \hline
LaAl & --1.01 & --0.96 \footnotemark[1] \\
LaAl$_2$ & --1.76 & {--1.58 }\footnotemark[1] \\
LaAl$_3$ & --2.13 & {--1.83 }\footnotemark[1] \\
La$_3$Al$_{11}$ & --6.51 & {--5.97 }\footnotemark[1] \\
LaGe & --1.97 & --1.92 \footnotemark[2] \\
La$_5$Ge$_3$ & --6.82 & {--6.96 }\footnotemark[2] \\
LaAl$_2$Ge$_2$ & --2.81 & N/A \\
\end{tabular}
\end{ruledtabular}
\footnotetext[1]{Reference ~\onlinecite{Enthalpy_Al_01}}
\footnotetext[2]{Reference ~\onlinecite{NIST-Al2O3}}
\footnotetext[3]{Reference ~\onlinecite{Enthalpy_Al_LAO_03}}
\end{table}

\begin{figure*}[t]
\includegraphics[width = 17 cm]{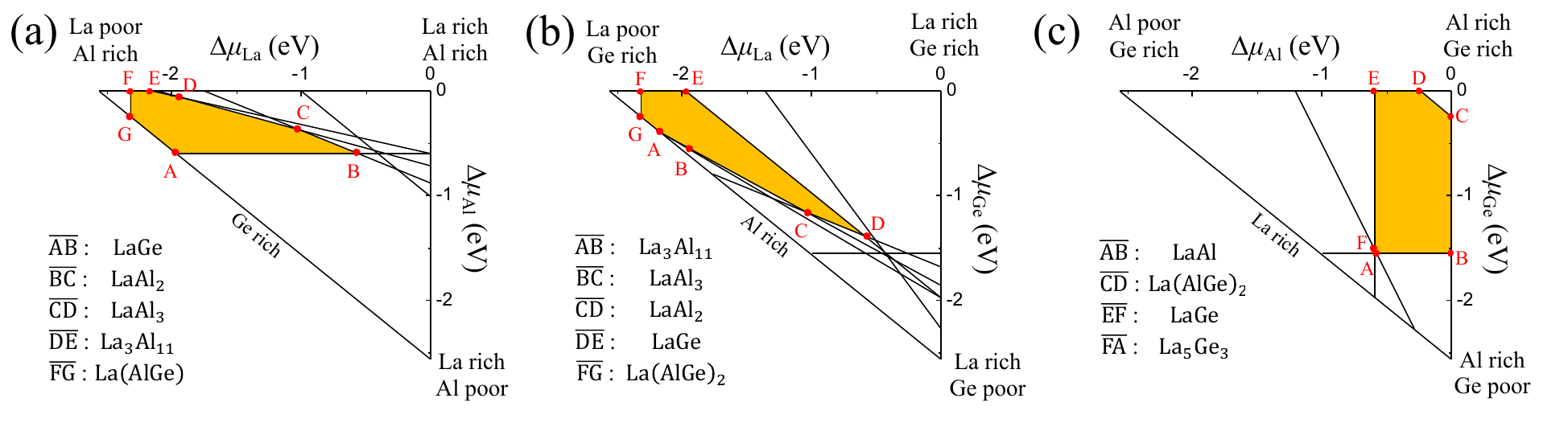}\\
\caption{\label{chempot} Allowed values of the atomic chemical potentials defining the stability of LaAlGe. The chemical potentials $\mu_\mathrm{La}$, $\mu_\mathrm{Al}$, and $\mu_\mathrm{Ge}$ are limited by the formation of the secondary phases as described in the text and Table. The filed region indicates the conditions that enable to growth LaAlGe. 
}
\end{figure*}

In this formalism the atomic chemical potentials $\mu_{i}$ are variable, yet restricted by the
formation of limiting phases containing La, Al, and Ge. The atomic chemical potentials $\mu_{i}$ must satisfy the stability condition of LaAlGe:
\begin{equation} \label{Eq3}
\   \mu_{\rm La} +  \mu_{\rm Al}+   \mu_{\rm Ge} = \Delta H _f {\rm (LaAlGe)},
\end{equation}
with $\Delta\mu_{\rm La}$ $\le$ 0, $\Delta\mu_{\rm Al}$ $\le$ 0, and $\Delta\mu_{\rm Ge}$ $\le$ 0. $\Delta H_f {\rm (LaAlGe)}$ is the formation enthalpy of perfect LaAlGe crystal. The chemical potentials $\mu_{\rm La}$, $\mu_{\rm Al}$, and $\mu_{\rm Ge}$ are further constrained by the formation of the secondary phases, such as LaAl, LaAl$_2$, LaAl$_3$, La$_3$Al$_{11}$, LaGe, La$_5$Ge$_3$,  and La(AlGe)$_2$ phases:
\begin{eqnarray} \label{Eq4}
\ \Delta  \mu_{\rm La} + \Delta \mu_{\rm Al} < \Delta H _f {\rm (LaAl)},
\end{eqnarray}
\begin{eqnarray} \label{Eq5}
\ \Delta \mu_{\rm La} + 2\Delta \mu_{\rm Al} < \Delta H _f {\rm (LaAl_2)},
\end{eqnarray}
\begin{eqnarray} \label{Eq6}
\ \Delta \mu_{\rm La} + 3\Delta \mu_{\rm Al} < \Delta H _f {\rm (LaAl_3)},
\end{eqnarray}
\begin{eqnarray} \label{Eq7}
\  \Delta \mu_{\rm La} + 11 \Delta \mu_{\rm Al} < \Delta H _f {\rm (LaAl_{11})},
\end{eqnarray}
\begin{eqnarray} \label{Eq8}
\ \Delta \mu_{\rm La} + \Delta \mu_{\rm Ge} < \Delta H _f {\rm (LaGe)},
\end{eqnarray}
\begin{eqnarray} \label{Eq9}
\ 5\Delta \mu_{\rm La} + 3\Delta \mu_{\rm Ge} < \Delta H _f {\rm (La_5Ge_3)},
\end{eqnarray}
\begin{eqnarray} \label{Eq10}
\ \Delta \mu_{\rm La} + 2\Delta \mu_{\rm Al} + 2\Delta \mu_{\rm Ge} < \Delta H _f {\rm (La(AlGe)_2)},
\end{eqnarray}
The calculated formation enthalpies of the phases are listed in Table \ref{enthalpy}.  We used the equations to define a region in which LaAlGe is thermodynamically stable (See the filled regions in Fig.~\ref{chempot}).

\begin{table}[b]
\caption{\label{param} Lattice constants, interatomic distances, and formation enthalpy of LaAlGe obtained using the HSE and the generalized-gradient-approximation (GGA) functional of Perdew, Burke, and Ernzerhof (PBE)~\cite{pbe}. The value in parenthesis is a number of the nearest neighboring bonds (distance). Experimental values are also shown for comparison.  }
\begin{ruledtabular}
\begin{tabular}{cccc}
%& \multicolumn{3}{c}{Cubic}   \\
%\cline{2-4}
     Property  & GGA & HSE &Experiment \\
    \hline
    $a$ (\AA)  & 4.380& 4.348 & \makecell{4.336\footnotemark[1]\\4.344\footnotemark[2]\\4.349\footnotemark[3]} \\
    $c$ (\AA)  & 14.876 & 14.736 & \makecell{14.828\footnotemark[1]\\14.812\footnotemark[2]\\14.829\footnotemark[3]} \\
     $d_{\rm La-Ge}$ (\AA)  & \makecell{3.303 ($\times$2) \\ 3.339 ($\times$4)} & \makecell{3.282 ($\times$2) \\ 3.309 ($\times$4)} & \makecell{3.280 ($\times$2)\footnotemark[1] \\ 3.310 ($\times$4)\footnotemark[1]}  \\
    $d_{\rm La-Al}$ (\AA)  & \makecell{3.310 ($\times$2) \\ 3.335 ($\times$4)} & \makecell{3.279 ($\times$2) \\ 3.311 ($\times$4)} & \makecell{3.300 ($\times$6)\footnotemark[1]}  \\
   $d_{\rm Al-Ge}$ (\AA)  & \makecell{2.484 \\ 2.515 ($\times$2)} & \makecell{2.454 \\ 2.498 ($\times$2)} & \makecell{2.498\footnotemark[1] \\ 2.470 ($\times$2)\footnotemark[1]}  \\
    $\Delta H^f$ (eV/f.u.)  & --2.138 & --2.562 & N/A\\
\end{tabular}
\end{ruledtabular}
\footnotetext[1]{Reference ~\onlinecite{Guloy_91}}
\footnotetext[2]{Reference ~\onlinecite{osti_1473922}}
\footnotetext[3]{Reference ~\onlinecite{Cao_PRB_22}}
\end{table}

%%%%%%%%%%%%%%%%%%%%%%%%%%%%%%%%%%%%%%%%%%%%%%%%%%%%%%
%%%%%%%%%%%%%%%%%%%%%%%%%%%%%%%%%%%%%%%%%%%%%%%%%%%%%%

\section{Results and discussion}
\label{sec:natdef}

%%%%%%%%%%%%%%%%%%%%%%%%%%%%%%%%%%%%%%%%%%%%%%%%%%%%%%
\subsection{Properties of pristine LaAlGe}
%%%%%%%%%%%%%%%%%%%%%%%%%%%%%%%%%%%%%%%%%%%%%%%%%%%%%%

\begin{figure}[]
\includegraphics[width =  8.8 cm]{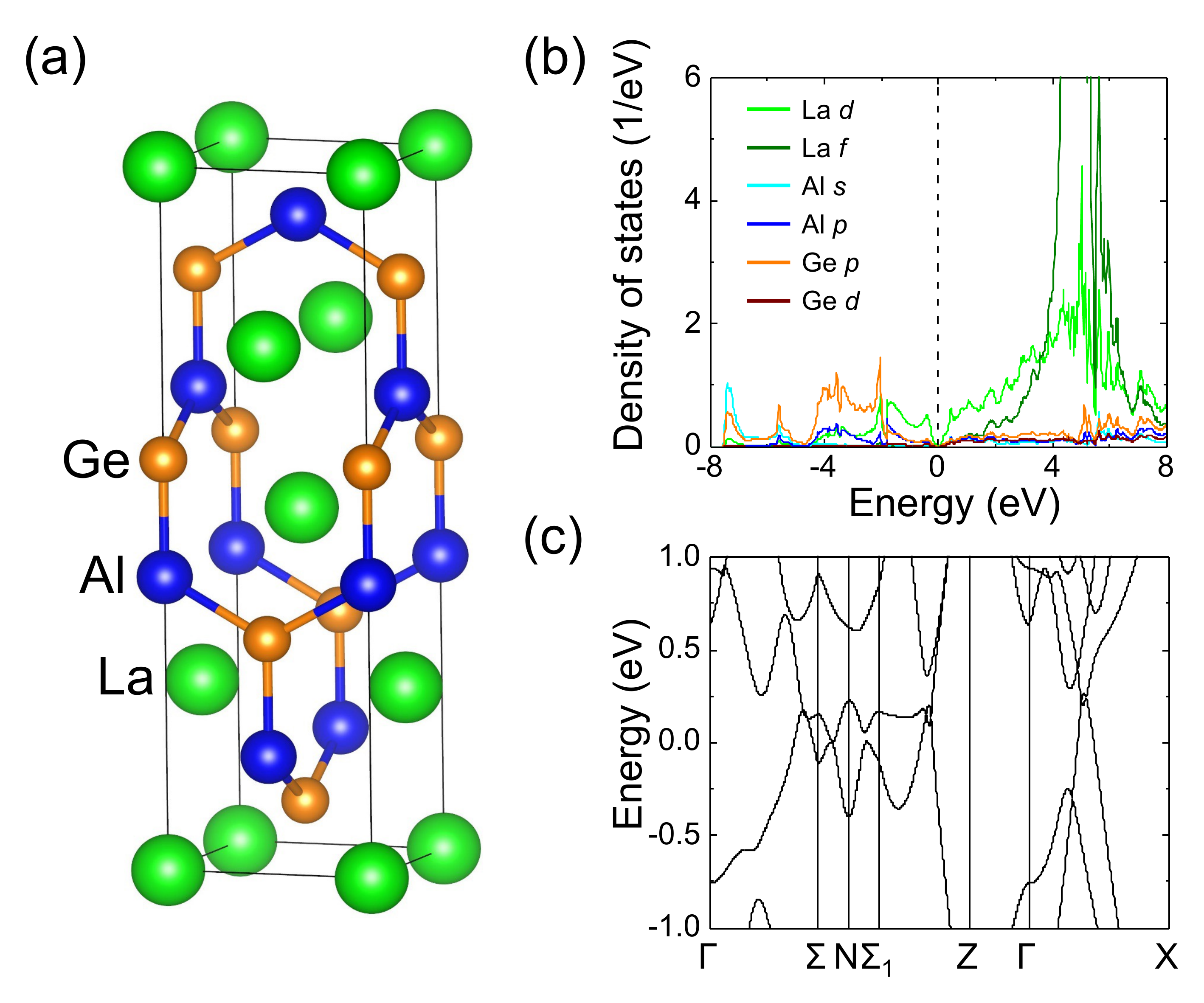}\\
\caption{\label{fig_crystal} (a) Crystal structure, (b) density of states, and (c) band structure of LaAlGe obtained using the HSE functional without the spin-orbit coupling.}
\end{figure}

 Prior to investigating native defects in LaAlGe, the structural and electronic properties of pristine LaAlGe were examined. Figure~\ref{fig_crystal} shows the conventional unit cell of LaAlGe and its band structure and density of states (DOS) obtained using the HSE functional. The calculated lattice parameters and formation enthalpy of LaAlGe agree well with experimental values (Table \ref{param})\cite{Howard_LatParam_00}. The electronic structure shows a semi-metallic feature which is consistent with the experiment. LaAlGe has a low DOS near the Fermi level with characteristics of La 5$d$. This indicates that as electrons are added or removed, the position of the Fermi level readily moves up or down. 
 
 %Most of the La 4$f$ states are found at $\sim$ 5 eV above the Fermi level, which is consistent with those in LaAlO3 (Reference) 

In order to validity the electronic structure obtained using the HSE functional,  LQSGW+DMFT calculations were performed to obtain the DOS of pristine LaAlGe which has been successfully utilized to address the electronic structure of correlated material systems, including actinide-and lanthanide-based compounds~\cite{siddiquee2022breakdown,kang2023dual,kang2022orbital,kang2022tunable,kang2023infinite,kang2019nio}. We find that overall features agree with the HSE results. An important observation is the position of the main peak of La 4$f$ state near 5 eV above the Fermi level, which agrees with the HSE but differs from the PBE-GGA result (See Fig. 2 in the supplementary material~\cite{supp}).

\begin{figure*}[]
	\includegraphics[width = 16 cm]{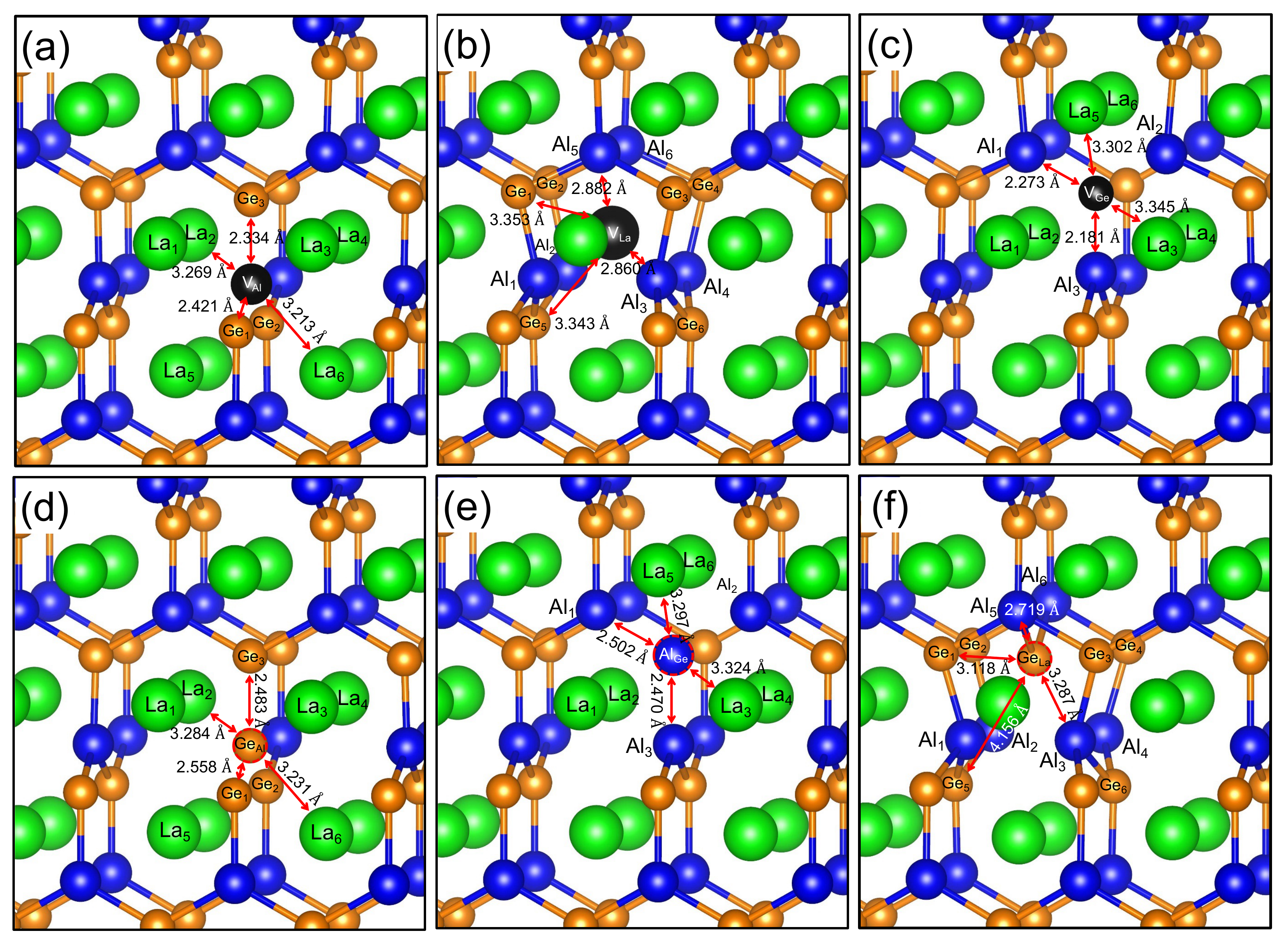}\\
	\caption{\label{fig_defect_structure} Local atomic structures of the lattice vacancies ((a) $V_{\rm Al}$, (b) $V_{\rm La}$, and (c) $V_{\rm Ge}$) and three antisite defects ((d) Ge$_{\rm Al}$, (e) Al$_{\rm Ge}$, and (f) Ge$_{\rm La}$), which are mostly likely to form in LaAlGe~\cite{VESTA_11}. Black sphere indicates the position of the vacant site, and the interatomic distances near the defects are shown. 
}
\end{figure*}

%%%%%%%%%%%%%%%%%%%%%%%%%%%%%%%%%%%%%%%%%%%%%%%%%%%%%%
\subsection{Vacancies}
%%%%%%%%%%%%%%%%%%%%%%%%%%%%%%%%%%%%%%%%%%%%%%%%%%%%%%

Figure~\ref{fig_defect_structure} shows the obtained local vacancy structures.
$V_{\rm Al}$ leads to significant inward relaxation of the surrounding three Ge atoms, with $V_{\rm Al}$--Ge distances of 2.421 {\AA} for $\rm Ge_{1}$ and $\rm Ge_{2}$, and 2.334 {\AA} for $\rm Ge_{3}$; these are --3.08 \% and --4.89 \% smaller than the  Al--Ge bond lengths in pristine LaAlGe, respectively (Table \ref{param}). Similarly, $V_{\rm Al}$--La distances are 3.269 {\AA} for $\rm La_{1}$ to $\rm La_{4}$, and 3.213 {\AA} for $\rm La_{5}$ and $\rm La_{6}$, which are --1.27 \% and --2.01 \% smaller than the  Al--La bond lengths in pristine LaAlGe. The calculated formation energies of all the native point defects in ternary LaAlGe for points indicated in the diagrams of Fig.~\ref{chempot} were evaluated (Table~\ref{formE}). Al vacancies have the lowest formation energies among the vacancies, even for Al-rich conditions (e.g., points D and E in Fig.~\ref{chempot} (c)). However, their formation energies for all the conditions considered here are higher than 4.5 eV, hence the formation of isolated vacancies during growth may be not likely.

\begin{table*}[]
\caption{\label{formE} Calculated formation energy of the vacancies and antisites in LaAlGe for points indicated  in the diagrams of Fig.~\ref{chempot}. Chemical potentials of $\Delta \mu_{\rm La}$ vs $\Delta \mu_{\rm Al}$, $\Delta \mu_{\rm La}$ vs $\Delta \mu_{\rm Al}$, $\Delta \mu_{\rm La}$ vs $\Delta \mu_{\rm Al}$ correspond to Figs.~\ref{chempot}(a), (b), and (c), respectively. }
\begin{ruledtabular}
\begin{tabular}{cccccccccc}
Chemical potential & \multicolumn{3}{c}{Vacancies} &  \multicolumn{6}{c}{Antisites} \\
\cline{2-4}   \cline{5-10}
$\Delta \mu_{\rm La}$ vs $\Delta \mu_{\rm Al}$   & $V_{\rm La}$ & $V_{\rm Al}$ & $V_{\rm Ge}$ & La$_{\rm Al}$ & La$_{\rm Ge}$ & Al$_{\rm La}$ & Al$_{\rm Ge}$ & Ge$_{\rm La}$ & Ge$_{\rm Al}$\\
    \hline
A & 7.40 & 4.53 & 8.12 & 3.63 & 8.43 & 3.61 & 2.64 & 1.69 & --1.38 \\
B & 8.80 & 4.53 & 6.73 & 2.23 & 5.64 & 5.01 & 1.24 & 4.48 & 0.02 \\
C & 8.35 & 4.75 & 6.95 & 2.90 & 6.31 & 4.33 & 1.24 & 3.81 & 0.02 \\
D & 7.43 & 5.06 & 7.56 & 4.13 & 7.84 & 3.10 & 1.55 & 2.27 & --0.29 \\
E & 7.20 & 5.12 & 7.73 & 4.42 & 8.24 & 2.82 & 1.65 & 1.88 & --0.39 \\
F & 7.05 & 5.12 & 7.88 & 4.57 & 8.53 & 2.67 & 1.80 & 1.58 & --0.54 \\
G & 7.05 & 4.88 & 8.12 & 4.33 & 8.78 & 2.91 & 2.29 & 1.34 & --1.03 \\
   \hline
$\Delta \mu_{\rm La}$ vs $\Delta \mu_{\rm Ge}$ & $V_{\rm La}$ & $V_{\rm Al}$ & $V_{\rm Ge}$ & La$_{\rm Al}$ & La$_{\rm Ge}$ & Al$_{\rm La}$ & Al$_{\rm Ge}$ & Ge$_{\rm La}$ & Ge$_{\rm Al}$\\
    \hline
A & 7.20 & 5.12 & 7.73 & 4.42 & 8.24 & 2.82 & 1.65 & 1.88 & --0.39 \\
B & 7.43 & 5.06 & 7.56 & 4.13 & 7.84 & 3.10 & 1.55 & 2.27 & --0.29 \\
C & 8.35 & 4.75 & 6.95 & 2.90 & 6.31 & 4.33 & 1.24 & 3.81 & 0.02 \\
D & 8.80 & 4.53 & 6.73 & 2.23 & 5.64 & 5.01 & 1.24 & 4.48 & 0.02 \\
E & 7.40 & 4.53 & 8.12 & 3.63 & 8.43 & 3.61 & 2.64 & 1.69 & --1.38 \\
F & 7.05 & 4.88 & 8.12 & 4.33 & 8.78 & 2.91 & 2.29 & 1.34 & --1.03 \\
G & 7.05 & 5.12 & 7.88 & 4.57 & 8.53 & 2.67 & 1.80 & 1.58 & --0.54 \\
   \hline
$\Delta \mu_{\rm Al}$ vs $\Delta \mu_{\rm Ge}$  & $V_{\rm La}$ & $V_{\rm Al}$ & $V_{\rm Ge}$ & La$_{\rm Al}$ & La$_{\rm Ge}$ & Al$_{\rm La}$ & Al$_{\rm Ge}$ & Ge$_{\rm La}$ & Ge$_{\rm Al}$\\
    \hline
A & 8.94 & 4.55 & 6.57 & 2.10 & 5.34 & 5.13 & 1.07 & 4.78 & 0.19 \\
B & 8.36 & 5.12 & 6.57 & 3.26 & 5.91 & 3.98 & 0.49 & 4.21 & 0.77 \\
C & 7.05 & 5.12 & 7.88 & 4.57 & 8.53 & 2.67 & 1.80 & 1.58 & --0.54 \\
D & 7.05 & 4.88 & 8.12 & 4.33 & 8.78 & 2.91 & 2.29 & 1.34 & --1.03 \\
E & 7.40 & 4.53 & 8.12 & 3.63 & 8.43 & 3.61 & 2.64 & 1.69 & --1.38 \\
F & 8.91 & 4.53 & 6.61 & 2.11 & 5.40 & 5.12 & 1.12 & 4.72 & 0.14 \\
\end{tabular}
\end{ruledtabular}
\end{table*}

$V_{\rm La}$ leads to insignificant outward relaxation of the surrounding six Ge atoms, with $V_{\rm La}$--Ge distances of 3.353 {\AA} for $\rm Ge_{1}$, $\rm Ge_{2}$, $\rm Ge_{3}$ and $\rm Ge_{4}$, and 3.343 {\AA} for $\rm Ge_{5}$ and $\rm Ge_{6}$; these are 1.33 \% and 1.86 \% larger than the La--Ge bond lengths in pristine LaAlGe, respectively. However, the surrounding six Al atoms are significantly inward relaxed. The $V_{\rm La}$--Al distances are 2.860 {\AA} (--13.62 \%) for $\rm Al_{1}$ to $\rm Al_{4}$ and 2.882 {\AA} (--12.11 \%) for $\rm Al_{1}$ and $\rm Al_{2}$.

Such significant inward relaxations are also observed for $V_{\rm Ge}$. The $V_{\rm Ge}$--Al distances of 2.273 {\AA} for $\rm Al_{1}$ and $\rm Al_{2}$, and 2.181 {\AA} for $\rm Al_{3}$ correspond to --9.01 \% and --11.12 \% shortening compared with the Ge--Al bond lengths in pristine LaAlGe, respectively. On the other hand, the distances with the surrounding six La atoms are relatively robust. The change in the Ge-La distances are 1.09 \% for $\rm La_{1}$, $\rm La_{2}$, $\rm La_{3}$ and $\rm La_{4}$, and --0.61 \% for $\rm La_{5}$ and $\rm La_{6}$.

In terms of energetics, the two vacancies $V_{\rm La}$ and $V_{\rm Ge}$ are not a concern in LaAlGe because of their quite high formation energies.  The lowest formation energies are 7.05 eV for $V_{\rm La}$ and 6.57 eV for $V_{\rm Ge}$, which indicates that their concentrations would be negligible.

%%%%%%%%%%%%%%%%%%%%%%%%%%%%%%%%%%%%%%%%%%%%%%%%%%%%%%
\subsection{Antisites}

Similar to the case of vacancies, one specific defect, the Ge$_{\rm Al}$ antisite (Fig.~\ref{fig_defect_structure}) has the lowest formation energy among the six types of antisites in LaAlGe, regardless of growth conditions (Table~\ref{formE}). The defect formation energies are very low for most growth conditions (even negative for several conditions), which indicates that the concentration of Ge$_{\rm Al}$ antisites should be quite high. Structurally, the  ${\rm Ge_{Al}}$--Ge distances are 2.558 {\AA} for $\rm Ge_{1}$ and $\rm Ge_{2}$, and 2.483 {\AA} for $\rm Ge_{3}$,  which are 2.40 \% and 1.18 \% larger than the Al--Ge bond lengths in pristine LaAlGe, respectively. The surrounding six La atoms are inward relaxed. The ${\rm Ge_{Al}}$--La distances are 3.284 {\AA} (--0.82 \%) for $\rm La_{1}$, $\rm La_{2}$, $\rm La_{3}$ and $\rm La_{4}$, and 3.231 {\AA} (--1.46 \%)  for $\rm La_{5}$ and $\rm La_{6}$.

The counterpart antisite of the Ge$_{\rm Al}$ antisite, the Al$_{\rm Ge}$ antisite also has relatively low formation energies with the mean value of 1.66 eV, and its formation energy becomes 0.49 eV under the Al-rich and Ge-poor condition (point B in Fig~\ref{chempot} (c)). Even though Ge$_{\rm Al}$ has the lowest formation energy in most conditions, the formation energy of Ge$_{\rm Al}$ is higher than that of Al$_{\rm Ge}$ at the point B (Al rich, Ge poor condition). The interatomic distances between Al$_{\rm Ge}$ and the nearest Al atoms increase a little, with the ${\rm Al_{Ge}}$--Al distances of 2.502 {\AA} (0.16 \%) for $\rm Al_{1}$ and $\rm Al_{2}$, and 2.470 {\AA} (0.65 \%) for $\rm Al_{3}$.  The Al$_{\rm Ge}$--La distances also slightly increases, with the ${\rm Al_{Ge}}$--La distances of 3.324 {\AA} (0.45 \%) for $\rm La_{1}$ to $\rm La_{4}$ and 3.297 {\AA} (0.46 \%) for $\rm La_{5}$ and $\rm La_{6}$.

Our results indicate that the formation of Ge$_{\rm La}$ antisite cannot be excluded when the LaAlGe samples are grown at La-poor or Ge-rich conditions since the defect formation energies become lower than 2 eV. However, Ge$_{\rm Al}$ would be dominant even for the growth conditions ($E^{f} ({\rm Ge_{Al}}) < 0$). %\textcolor{red}{Structurally, ${\rm Ge_{La}}$ antisite is posited near by two Al atoms ($\rm Al_{5}$ and $\rm Al_{6}$), shows significant atomic relaxation compared to aforementioned antisites. ${\rm Ge_{La}}$--Ge distances of 3.118 {\AA} (--5.77 \%) for $\rm Ge_{1}$, $\rm Ge_{2}$, $\rm Ge_{3}$ and $\rm Ge_{4}$, and 4.156 {\AA} (26.63 \%) for $\rm Ge_{5}$ and $\rm Ge_{6}$. On the other hand, the ${\rm Ge_{La}}$--Al distances of 3.287 {\AA} for $\rm Al_{1}$, $\rm Al_{2}$, $\rm Al_{3}$ and $\rm Al_{4}$, and 2.719 {\AA} for $\rm Al_{5}$ and $\rm Al_{6}$ correspond to --0.72 \% and --17.08 \% shortening compared with the La--Al bond lengths in pristine LaAlGe, respectively.}

On the other hand, the other antisites may be unlikely to form because of their very high formation energies. Especially, the formation energies of La$_{\rm Ge}$ are in the range of 5 to 9 eV, which is huge. Presumably, these defects may be able to form in LaAlGe only if growth techniques operated far from thermal equilibrium (e.g., ion implantation) are used to obtain the specimens or to introduce the defects.

%%%%%%%%%%%%%%%%%%%%%%%%%%%%%%%%%%%%%%%%%%%%%%%%%%%%%%
\subsection{Impact of defects on the properties of LaAlGe }
%\label{sec:align}
%%%%%%%%%%%%%%%%%%%%%%%%%%%%%%%%%%%%%%%%%%%%%%%%%%%%%%

\begin{figure}[]
\includegraphics[width = 8.5 cm]{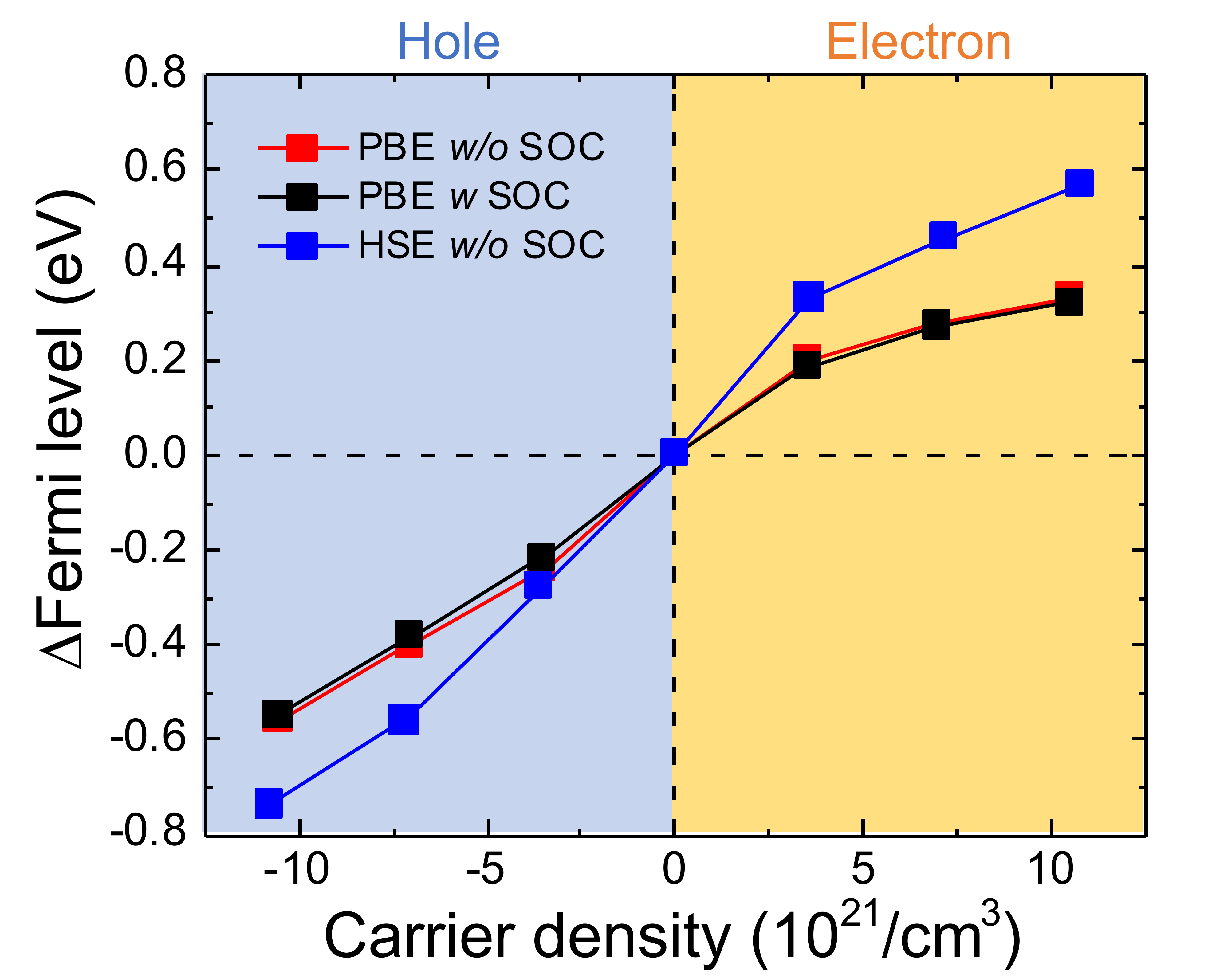}\\
\caption{\label{fig_Fermi} Calculated relative position of the Fermi level as a function of carrier concentration in LaAlGe. The Fermi level in the pristine LaAlGe is used as reference. The blue and orange regions denote LaAlGe doped by removing and adding electron, respectively. The results obtained using GGA-PBE with and without the spin-orbit coupling effect are shown for comparison.}
\end{figure}

Finally, we discussed the impact of native defects on the properties of LaAlGe. As mentioned in the Introduction section, growing LaAlGe single crystal is typically grown out of Al flux because of its low melting point due to the volatile nature of Al. The present study supports the requirement of an Al-rich environment to obtain high-quality LaAlGe crystal, which will affect the physics of Weyl fermion in LaAlGe. Our results show that Ge$_{\rm Al}$ is dominant defect and its formation cannot be unavoidable but the growth under Al-rich condition increases its formation energy which could relatively hinder the Ge$_{\rm Al}$ formation (e.g., point B in Fig.~\ref{chempot} (c)). This may help to obtain a better crystal quality. In addition, the counterpart defect, Al$_{\rm Ge}$ could help to get it. The formation energy of Al$_{\rm Ge}$ decreases under Al-rich condition, hence the defect becomes more likely to form then partly compensate electron carriers released from Ge$_{\rm Al}$.

When the two defects form with moderate concentrations, there may exist an interaction between them.  Thus, the binding energy of  a complex defect,  Ge$_{\rm Al}$--Al$_{\rm Ge}$, which is consist of Ge$_{\rm Al}$ and Al$_{\rm Ge}$ antisites, was evaluated using $E_{\rm bind} [({\rm Ge_{Al}}-{\rm Al_{Ge}})] =   E^f ({\rm Ge_{Al}}) + E^f ({\rm Al_{Ge}}) -E^f [({\rm Ge_{Al}}-{\rm Al_{Ge}})]$. The computed value is positive but low ($E_{\rm bind} [({\rm Ge_{Al}}-{\rm Al_{Ge}}$) = 0.52 eV). Therefore, the complex would occur during cool down because complex formation becomes energetically advantageous but can relatively easily dissociate. Note that the formation energies do not depend on the atomic chemical potentials, hence the values are constant regardless of growth conditions considered here.

Moving to electronic property, LaAlGe may have a large concentration of electron carriers, since the formation of Ge$_{\rm Al}$ during growth is very likely at high temperatures. Ge$_{\rm Al}$ can release one electron in the sample when the Ge$^{4+}$ and Al$^{3+}$ oxidation states assumed. Indeed, the measured electron concentrations in an experiment are 1.08 $\times$ 10$^{21}$ cm$^{-3}$ at 10 K and 1.06 $\times$ 10$^{21}$ cm$^{-3}$ at 20 K with growth temperature $T$ = 700 $^\circ$C \cite{2020_LAG_concen_APL}. Using the equation for the Ge$_{\rm Al}$ concentration ($c = N_{\rm sites} ~exp(-E^f  / k_{B}T)$,  where $N_{\rm sites}$ is the density of sites Al sites,  $E^f ({\rm Ge_{Al}})$ corresponds to 2.2 meV and 4.5 meV, which are close to the calculated values. Such electron concentrations in LaAlGe thin films and single crystals place the Fermi level at significantly higher positions compared to the expected value in the pristine materials, as shown in Fig.~\ref{fig_Fermi}. Our results indicate that the chemical potential can be pushed away from the Fermi level position due to the intrinsic electron doping by native defect, hence causing the unexpected experimental observation, different from the other materials of the $Ln$AlX family.

These results suggest a strategy for tuning the electronic chemical potential in these materials.  Since Ge$_{\rm Al}$ is the main concern, increasing the counterpart Al$_{\rm Ge}$ is the most straightforward to put the electronic chemical potential position near the Fermi level in perfect LaAlGe crystal, by controlling the growth environment close to the condition providing the formation energies of the two antisites similar such as point B in Fig.~\ref{chempot} (c) ($\Delta \mu _{\rm Al}$ = 0 eV,  $\Delta \mu _{\rm Ge}$ = --1.55 eV, and $\Delta \mu _{\rm La}$ = --1.01 eV). At the growth condition,  Ge$_{\rm Al}$ and Al$_{\rm Ge}$ are most likely to form with similar formation energies, and therefore, the most electron carriers would be compensated by holes towards the perfect LaAlGe, bringing back the Weyl physics. 

%%%%%%%%%%%%%%%%%%%%%%%%%%%%%%%%%%%%%%%%%%%%%%%%%%%%%%
%%%%%%%%%%%%%%%%%%%%%%%%%%%%%%%%%%%%%%%%%%%%%%%%%%%%%%

\section{Summary}

We have investigated native defects in the Weyl semimetal LaAlGe using hybrid density functional calculations by analyzing the defect formation energies. We find that vacancies are not likely to form during growth, whereas Ge- and Al-related antisite defects are likely to form. This indicates that the antisite defects can play a role in the growing high-quality crystal, which will affect the Weyl physics of LaAlGe by shifting the position of the electronic chemical potential. Especially, Ge$_{\rm Al}$ might provide high electron density to the samples due to its low formation energy for all the atomic chemical potentials considered here. The formation of Al$_{\rm Ge}$ is also comparable with specific growth conditions. The estimated binding energy shows Ge$_{\rm Al}$ and Al$_{\rm Ge}$ would prefer to bond and form a complex defect, but the complex may easily dissociate at high temperatures. Our results propose that the position of the chemical potential can be tuned by putting the atomic chemical potentials to the condition resulting in the formation of the two antisites are comparable, and therefore it might shift back to the Fermi level where the Weyl nodes are located.

\begin{acknowledgments}
This work was supported by Inha University Research Grant (INHA-68917). We also acknowledge the use of the KISTI supercomputing center (grant no. KSC-2023-CRE-0421). 
\end{acknowledgments}

%\textcolor{red}{**]]}

\bibliography{Choi_LAG}

\end{document}